\documentstyle[aps,pra,eqsecnum]{revtex}

\begin{document}
\title{Decoherence-Free Subspaces for Multiple-Qubit Errors: (I)
Characterization}
\author{Daniel A. Lidar,$^{1}$ Dave Bacon,$^{1,2}$ Julia
Kempe$^{1,3,4}$ and K.B. Whaley$^{1}$}
\address{Departments of Chemistry$^{1}$, Physics$^{2}$ and Mathematics$^{3}$,\\
University of California, Berkeley, CA 94720\\
\'{E}cole Nationale Superieure des T\'{e}l\'{e}communications,\\
Paris, France$^4$}
\date{\today}
\maketitle

\begin{abstract}
Coherence in an open quantum system is degraded through its interaction with
a bath. This decoherence can be avoided by restricting the dynamics of the
system to special decoherence-free subspaces. These subspaces are usually
constructed under the assumption of spatially symmetric system-bath
coupling. Here we show that decoherence-free subspaces may appear without
spatial symmetry. Instead, we consider a model of system-bath interactions
in which to first order only multiple-qubit coupling to the bath is present,
with single-qubit system-bath coupling absent. We derive necessary and
sufficient conditions for the appearance of decoherence-free states in this
model, and give a number of examples. In a sequel paper we show how to
perform universal and fault tolerant quantum computation on the
decoherence-free subspaces considered in this paper.

\end{abstract}

\section{Introduction}

Quantum information must be protected against the detrimental effects of
decoherence \cite{Lo:book,Williams:book98}. To this end Decoherence-Free
Subspaces (DFSs) \cite
{Duan:98,Duan:98c,Zanardi:97c,Zanardi:97a,Zanardi:98a,Lidar:PRL98,Lidar:PRL99,Bacon:99}
have recently been proposed, alongside Quantum Error Correcting Codes
(QECCs) \cite{Shor:95,Steane:96a,Gottesman:97,Knill:97b} and ``dynamical
decoupling'' and symmetrization schemes \cite
{Viola:98,Viola:99,Duan:98e,Zanardi:98b}. A DFS is a ``quiet corner'' of the
system's Hilbert space, where the evolution is decoupled from the bath and
thus is entirely unitary. DFSs are a special class of (fully degenerate)
QECCs \cite{Lidar:PRL99}, so in order to properly distinguish between DFSs
and all other QECCs we note that DFSs are {\em passive} codes, in that the
information encoded in them may not require any active stabilization
procedures \cite{Bacon:99a,Kempe:00}. All other QECCs, in contrast, always
involve an {\em active} error-detection/correction process. Examples of DFSs
have so far focused almost exclusively on the presence of a {\em permutation
symmetry} of some sort in the system-bath coupling. The most often used
example is that of \ ``collective decoherence'' \cite
{Duan:98,Duan:98c,Zanardi:97c,Lidar:PRL98,Lidar:PRA99Exchange}, where the
bath couples in an identical fashion to all qubits, implying that all qubits
undergo the {\em same} error. In this case four physical qubits suffice to
encode a logical qubit against any collective error, and the code efficiency
(number of encoded per physical qubits) approaches unity asymptotically \cite
{Zanardi:97c}. It was shown that the requirement of an exact symmetry can be
lifted by allowing for a symmetry-breaking perturbation, without spoiling
the DFS property significantly \cite{Lidar:PRL98,Bacon:99}. Moreover, by
concatenation with an active QECC, a symmetry-broken DFS can be stabilized
completely \cite{Lidar:PRL99}. While these results indicate that a small
departure from the exact symmetry condition for the system-bath coupling is
admissible, they leave unanswered the question of whether a DFS may exist
when no assumptions are made regarding the spatial symmetry of this coupling.

In this paper, the first out of two, it will be shown that under conditions
which do not relate to a spatially symmetric system-bath coupling, DFSs may
still exist. This result is exact, i.e., it is not of a perturbative nature
as in Refs.~\cite{Lidar:PRL98,Lidar:PRL99,Bacon:99}. Instead, it relies on
the assumption that errors affecting single qubits are absent, and to lowest
order only multiple-qubit errors are possible instead. Formally, the
condition is that the qubit register is not affected by the full Pauli group
of errors, but only by a subgroup thereof. One may then proceed to find DFSs
with respect to this subgroup. The interesting class of system-bath
interaction Hamiltonians which allow for such processes generally involve
only multiple-qubit operators. Relevant physical systems are therefore those
where the bath can only couple to multiple system excitations as is the case
for decoherence due to dipolar coupling, e.g., in NMR \cite{Slichter:96}.
Another interesting class of examples are composite particles, such as
bi-excitons in quantum dots/wells \cite{Chemla:98}, or Cooper pairs in
superconductors \cite{March:85vol2}.

The structure of this paper is as follows. In section \ref{Hamiltonians} we
briefly review the structure of Hamiltonians pertinent to systems that may
function as quantum computers, coupled to a decohering environment. Using
these Hamiltonians, we recall in section \ref{Evolution} the derivation of
the operator sum representation evolution equation for the system density
matrix. We show in particular that for a qubit system, the evolution can be
expressed entirely in terms of linear combinations of tensor products of
Pauli matrices. We then use this in section \ref{DFS} to derive the DFS
condition under the assumption that decoherence is the result of a subgroup
of the Pauli group. In section \ref{Examples} we illustrate our general
analysis with some examples, and find decoherence-free states for a number
of subgroups. We derive the dimension of these DFSs in section~\ref
{dimension}. Conclusions and a summary are presented in section \ref{Summary}. Finally, some important properties of the Pauli group are summarized in
Appendix \ref{app1}, and some examples of ``non-generic'' DFSs are presented
in Appendix \ref{app2}. We show in a sequel paper \cite{Lidar:00b} how to
perform universal fault tolerant quantum computation using at most 2-body
Hamiltonians on the DFSs derived here.

\section{Structure of the Hamiltonian for a Universal Quantum Computer
Coupled to a Bath}

\label{Hamiltonians}

This section provides a brief review of the structure of Hamiltonians
relevant for a qubit system allowing for universal quantum computation and
coupled to a decohering bath.

The dynamics of a quantum system $S$ coupled to a bath $B$ (which together
form a closed system) evolves unitarily under the combined Hamiltonian

\begin{equation}
{\bf H}={\bf H}_{S}{\bf \otimes }{\bf I}_{B}+{\bf I}_{S}{\bf \otimes H}_{B}+ 
{\bf H}_{I},
\end{equation}
where ${\bf H}_{S}$, ${\bf H}_{B}$ and ${\bf H}_{I}$ are the system, bath
and interaction Hamiltonians, respectively; ${\bf I}$ is the identity
operator. Let ${\bf \sigma }_{i}^{\alpha }$ denote the $\alpha ^{{\rm th}}$
Pauli matrix, $\alpha =\{0,x,y,z\}$, acting on qubit $i$. The $2\times 2$
identity matrix is denoted ${\bf \sigma }_{i}^{0}$. For $K$ qubits the
components of ${\bf H}$ can often be written as follows: 
\begin{equation}
{\bf H}_{S}=\sum_{i=1}^{K}\sum_{\alpha =x,z}\varepsilon _{i}^{\alpha }{\bf 
\sigma }_{i}^{\alpha }+\sum_{i\neq j}^{K}J_{ij}{\bf \sigma }_{i}^{+}{\bf 
\sigma }_{j}^{-}+{\rm h.c.},
\end{equation}
where ${\bf \sigma }_{i}^{\pm }=\left( {\bf \sigma }_{i}^{x}\mp i{\bf \sigma 
}_{i}^{y}\right) /2$. The first sum contains the qubit energies ($
\varepsilon _{i}^{z}$) and tunneling elements ($\varepsilon _{i}^{x}$) \cite
{Leggett:87}, and the second sum expresses tunneling between sites $i$ and $
j $. Other forms are also possible, e.g., as in an anisotropic dipolar
medium such as in solid state NMR \cite{Slichter:96}, where one would
typically encounter an Ising $J_{ij}^{z}{\bf \sigma }_{i}^{z}{\bf \sigma }
_{j}^{z}$ term. A Hamiltonian of the form above is sufficiently general to
allow for universal quantum computation by satisfying the following two
requirements \cite{DiVincenzo:95,Lloyd:95,Barenco:95a}: (i) Arbitrary
single-qubit operations are made possible by the presence of ${\bf \sigma }
_{i}^{x}$, which allows for the implementation of a continuous $SU(2)$
rotation in the $i^{{\rm th}}$ qubit Hilbert space, while the ${\bf \sigma }
_{i}^{z}$ term allows for the introduction of an arbitrary phase-shift
between the $|0\rangle $ and $|1\rangle $ states. When ${\bf \sigma }_{i}^{x}
$ and ${\bf \sigma }_{i}^{z}$ are exponentiated, they can be combined, using
the Lie sum and product formulae \cite{Bhatia} 
\begin{eqnarray}
\lim_{n\rightarrow \infty }\left( e^{i\alpha A/n}e^{i\beta B/n}\right) ^{n}&
=&e^{i(\alpha A+\beta B)}  \nonumber \\
\lim_{n\rightarrow \infty }\left( e^{iA/\sqrt{n}}e^{iB/\sqrt{n}}e^{-iA/\sqrt{
n}}e^{-iB/\sqrt{n}}\right) ^{n} &=&e^{[A,B]},  \label{eq:Lie}
\end{eqnarray}
to close the Lie algebra $su(2)$, and thus to construct any evolution in the
Lie group $SU(2)$ of all possible operations on a single qubit \cite
{Lloyd:95}. (ii) The second ingredient needed for universal quantum
computation is the controlled-not ({\sc CNOT}) gate, which is made possible
through the ability to implement each of the (nearest neighbor) $J_{ij}{\
\sigma }_{i}^{+}{\sigma }_{j}^{-}+{\rm h.c.}$ terms. When exponentiated,
such a term yields:

\[
{\bf U}_{\theta }=\left( 
\begin{array}{cccc}
1 & 0 & 0 & 0 \\ 
0 & \cos \theta  & i\sin \theta  & 0 \\ 
0 & i\sin \theta  & \cos \theta  & 0 \\ 
0 & 0 & 0 & 1
\end{array}
\right) 
\]
with $\theta \propto J_{ij}t$. For $\theta =\pi /4$ this is (up to a phase)
the ``square-root-swap'' operation, which when combined with single-qubit
rotations allows for the implementation of {\sc CNOT}. Alternatively, a $
J_{ij}^{z}{\bf \sigma }_{i}^{z}{\bf \sigma }_{j}^{z}$ term alone is
sufficient, since it can be used to implement a controlled-phase-shift, as
is done routinely in NMR \cite{Gershenfeld:97}. It is important to emphasize
that the universal gates construction just described is but one of many
different ways to achieve universal quantum computation. In fact, universal
gates implementing logic operations directly on physical qubits (as above)
are generally inappropriate for the purpose of {\em fault tolerant}
computation \cite{Preskill:99}. We consider a different gate construction in
the sequel paper \cite{Lidar:00b}, operating instead on ``encoded'' qubits,
which can be used to implement universal fault tolerant quantum computation.
For a useful survey of different universal and fault tolerant sets of gates
see Ref.~\cite{Boykin:99}.

The bath Hamiltonian can be written as 
\begin{eqnarray}
{\bf H}_{B}=\sum_{k}\omega _{k}{\bf B}_{k}
\end{eqnarray}
where, e.g., for the spin-boson Hamiltonian, ${\bf B}_{k}={\bf b}
_{k}^\dagger {\bf b}_{k}$ \cite{Leggett:87}, and ${\bf b}_{k}^{\dagger }$, $
{\bf b}_{k} $ are respectively creation and annihilation operators of bath
mode $k$.

Finally, the system-bath interaction Hamiltonian is 
\begin{eqnarray}
{\bf H}_{I}=\sum_{i=1}^{K}\sum_{\alpha =+,-,z}\sum_{k}g_{ik}^{\alpha }\sigma
_{i}^{\alpha }\otimes {\tilde{{\bf B}}}_{k}^{\alpha },
\label{eq:H_I1}
\end{eqnarray}
where $g_{ik}^{\alpha }$ is a coupling coefficient. In the spin-boson model
one would have ${\tilde{{\bf B}}}_{k}^{+}={\bf b}_{k}$, ${\tilde{{\bf B}}}
_{k}^{-}={\bf b}_{k}^{\dagger }$ and ${\tilde{{\bf B}}}_{k}^{z}={\bf b}
_{k}^{\dagger }+{\bf b}_{k}$. Thus $\sigma _{i}^{\pm }\otimes {\tilde{{\bf B}
}}_{k}^{\pm }$ expresses a dissipative coupling (in which energy is
exchanged between system and environment), and $\sigma _{i}^{z}\otimes {\ 
\tilde{{\bf B}}}_{k}^{z}$ corresponds to a phase damping process (in which
the environment randomizes the system phases, e.g., through elastic
collisions).

An interesting limiting case arises when the coupling constants are
independent of the qubit index:\ $g_{ik}^{\alpha }\equiv g_{k}^{\alpha }$.
This situation, known as ``collective decoherence'', arises when there is
full permutational symmetry of qubit positions, and implies the existence of
a large DFS \cite{Zanardi:97c,Lidar:PRA99Exchange}. Defining collective
system operators $S^{\alpha }\equiv \sum_{i=1}^{K}\sigma _{i}^{\alpha }$,
one can then express the interaction Hamiltonian in greatly simplified form
as 
\[
{\bf H}_{I}^{{\rm coll.}}=\sum_{\alpha =+,-,z}S^{\alpha }\otimes \left(
\sum_{k}g_{k}^{\alpha }{\tilde{{\bf B}}}_{k}^{\alpha }\right) .
\]
A case of intermediate symmetry arises when the coupling constants are equal
not over the entire qubit register but rather only over finite clusters $
j=1..C$. One can then define cluster system operators $S_{j}^{\alpha }\equiv
\sum_{i_{j}=1}^{K_{j}}\sigma _{i_{j}}^{\alpha },$ where $K_{j}$ is the
number of qubits in cluster $j$. The interaction Hamiltonian becomes 
\[
{\bf H}_{I}^{{\rm clus.}}=\sum_{j=1}^{C}\sum_{\alpha =+,-,z}S_{j}^{\alpha
}\otimes \left( \sum_{k}g_{jk}^{\alpha }{\tilde{{\bf B}}}_{k}^{\alpha
}\right) .
\]
In this case too, DFSs can be found. The point we wish to emphasize
presently is that the underlying assumption in cluster decoherence is that
of {\em spatial symmetry} in the system-bath coupling. This is to be
contrasted with the decoherence models studied in this paper, where DFSs
will be shown to arise without the need for spatial symmetry.

Returning to the general case, ${\bf H}_{I}$ can be rewritten as 
\begin{eqnarray}
{\bf H}_{I}=\sum_{i=1}^{K}\sum_{\alpha =x,y,z}\sum_{k}\sigma _{i}^{\alpha
}\otimes {\bf B}_{ik}^{\alpha },  \label{eq:H_I}
\end{eqnarray}
where ${\bf B}_{ik}^{z}\equiv {\tilde{{\bf B}}}_{k}^{z}$ and ${\bf B}
_{ik}^{x}$ ,${\bf B}_{ik}^{y}$ are appropriate linear combinations of ${\ 
\tilde{{\bf B}}}_{k}^{+}$ and ${\tilde{{\bf B}}}_{k}^{-}$: 
\begin{eqnarray}
{\bf B}_{ik}^{x} &=&\frac{1}{2}\left( g_{ik}^{-}{\tilde{{\bf B}}}
_{k}^{-}+g_{ik}^{+}{\tilde{{\bf B}}}_{k}^{+}\right) \\
{\bf B}_{ik}^{y} &=&\frac{i}{2}\left( g_{ik}^{-}{\tilde{{\bf B}}}
_{k}^{-}-g_{ik}^{+}{\tilde{{\bf B}}}_{k}^{+}\right)
\end{eqnarray}
The qubit-coupling term in ${\bf H}_{S}$ can also be expressed entirely in
terms of $\sigma _{i}^{\alpha }$, where $\alpha =x,y$ or $z$. Thus all
system components of the Hamiltonian ${\bf H}$ can be expressed in terms of
tensor products of the single qubit {\em Pauli matrices}.

\section{Time Evolution of the Density Matrix}

\label{Evolution}

The purpose of this section is to show that the evolution of the density
matrix of an open system can be expanded in terms of tensor products of the
Pauli matrices (the Pauli group), and that this follows from the structure
of the Hamiltonians assumed above for a qubit register. This result is
obvious from a formal mathematical point of view (since the elements of the
Pauli group of order $K$ form a complete orthogonal set for the $2^{K}\times
2^{K}$ matrices) \cite{Chuang:97c}, so that the reader for whom this type of
argument is satisfactory may safely skip ahead to the next section. We
present the derivation of this result here in order to motivate the
appearance of the multiple-qubit errors that are the subject of this paper.

We first transform to the interaction picture \cite{Gardiner:book} defined
by the system and bath Hamiltonians: 
\begin{eqnarray}
{\bf H}\rightarrow {\bf H}(t)={\bf U}_{SB}(t){\bf HU}_{SB}^{\dagger }(t)= 
{\bf H}_{S}{\bf \otimes }{\bf I}_{B}+{\bf I}_{S}{\bf \otimes H}_{B}+{\bf H}
_{I}(t)
\end{eqnarray}
where 
\begin{eqnarray*}
{\bf U}_{SB}(t) &=&\exp \left[ -\left( {\bf H}_{S}{\bf \otimes }{\bf I}_{B}+ 
{\bf I}_{S}{\bf \otimes H}_{B}\right) it/\hbar \right] \\
&=&\exp \left[ -it{\bf H}_{S}/\hbar \right] {\bf \otimes }\exp \left[ -it 
{\bf H}_{B}/\hbar \right] ={\bf U}_{S}(t){\bf \otimes U}_{B}(t).
\end{eqnarray*}
Because the system and bath operators commute, the interaction picture
interaction Hamiltonian can be written as: 
\begin{eqnarray}
{\bf H}_{I}(t)={\bf U}_{SB}(t){\bf H}_{I}{\bf U}_{SB}^{\dagger
}(t)=\sum_{i=1}^{K}\sum_{\alpha =x,y,z}\sum_{k}\sigma _{i}^{\alpha
}(t)\otimes {\bf B}_{ik}^{\alpha }(t),
\end{eqnarray}
where 
\begin{eqnarray}
\sigma _{i}^{\alpha }(t) &=&{\bf U}_{S}(t)\sigma _{i}^{\alpha }{\bf U}
_{S}^{\dagger }(t)=\sum_{j,\beta }\lambda _{ij}^{\alpha \beta }(t)\sigma
_{j}^{\beta },  \nonumber \\
{\bf B}_{ik}^{\alpha }(t) &=&{\bf U}_{B}(t){\bf B}_{ik}^{\alpha }{\bf U}
_{B}^{\dagger }(t)  \label{eq:sigma_I}
\end{eqnarray}
(see, e.g., Ref.~\cite{Gardiner:book} for an explicit calculation of the $
\lambda _{ij}^{\alpha \beta }(t)$ for some examples). The system-bath
density matrix is transformed accordingly from the Schr\"{o}dinger into the
interaction picture (denoted by a prime): 
\begin{eqnarray}
\rho _{SB}(t)\longmapsto \rho _{SB}^{\prime }(t)={\bf U}_{SB}^{\dagger
}(t)\rho _{SB}(t){\bf U}_{SB}(t),
\end{eqnarray}
and the full dynamics is 
\begin{eqnarray}
\rho _{SB}^{\prime }(t)={\bf U}(t)\rho _{SB}^{\prime }(0){\bf U}^{\dagger
}(t),
\end{eqnarray}
where 
\begin{eqnarray}
{\bf U}(t)={\rm T}\exp \left[ -\frac{i}{\hbar }\int_{0}^{t}{\bf H}
_{I}(\tau)d\tau \right]
\end{eqnarray}
and ${\rm T}$ is the Dyson time-ordering operator (defined explicitly
below). From now on we work in the interaction picture only, so for
notational simplicity the prime is dropped from the density matrices. At $
t=0 $ the Schr\"{o}dinger and interaction pictures coincide. Thus, assuming
that system and bath are initially decoupled so that $\rho _{SB}(0)=\rho
(0)\otimes \rho _{B}(0)$, where $\rho $ and $\rho _{B}$ are, respectively,
the system and bath density matrices, the system dynamics is described by
the reduced density matrix:

\[
\rho (0)\longmapsto \rho (t)=\text{{\rm Tr}}_{B}[{\bf U}(t)(\rho (0)\otimes
\rho _{B}(0)){\bf U}^{\dagger }(t)].
\]
Here {\rm Tr}$_{B}$ is the partial trace over the bath. By using a spectral
decomposition for the bath, $\rho _{B}(0)=\sum_{\nu }p_{\nu }|\nu \rangle
\langle \nu |$,\footnote{
For a bath in thermal equilibrium, $|\nu \rangle $ would be an energy
eigenstate with energy $E_{\nu }$, and $p_{\nu }=\exp (-\beta E_{\nu })/Z$,
where $\beta $ is the inverse temperature and $Z={\rm Tr}[\exp (-\beta {\bf H
}_{B})]$ is the canonical partition function.} this can be rewritten in the
``operator sum representation'' \cite
{Bacon:99,Kraus:83,Schumacher:96a,Peres:99}:

\begin{eqnarray}
\rho (t)=\sum_{d}{\bf A}_{d}(t)\,\rho (0)\,{\bf A}_{d}^{\dagger }(t)
\label{eq:OSR}
\end{eqnarray}
where

\begin{eqnarray}
{\bf A}_{d}(t)=\sqrt{p_{\nu }}\langle \mu |{\bf U}(t)|\nu \rangle \;;\qquad
d=(\mu ,\nu )
\label{eq:Amunu}
\end{eqnarray}
Also, by unitarity of ${\bf U}$, one derives the normalization condition,

\begin{eqnarray}
\sum_{d}{\bf A}_{d}^{\dagger }{\bf A}_{d}={\bf I}_S  \label{eq:OSRnorm}
\end{eqnarray}
which guarantees preservation of the trace of $\rho $:

\begin{eqnarray}
{\rm Tr}[\rho (t)]={\rm Tr}[\sum_{d}{\bf A}_{d}\,\rho (0)\,{\bf A}
_{d}^{\dagger }]={\rm Tr}[\rho (0)\sum_{d}{\bf A}_{d}^{\dagger }{\bf A}
_{d}]= {\rm Tr}[\rho (0)].
\end{eqnarray}
The $\{{\bf A}_{d}\}$, called the {\em Kraus operators}, belong to the
(Banach, or Hilbert-Schmidt) space ${\cal B}({\cal H})$ of bounded operators
acting on the system Hilbert space, and for $K$ qubits are represented by $
2^{K}\times 2^{K}$ matrices, just like $\rho $.\footnote{
See, however, Ref.~\cite{Peres:99} for a discussion of Kraus operators
represented by non-square matrices.}

Consider now a formal Taylor expansion of the propagator: 
\begin{eqnarray}
{\bf U}(t) &=&\sum_{n=0}^{\infty }\frac{\left( -i\right) ^{n}}{n!}{\rm T}
\left( \int^{t}{\bf H}_{I}(\tau )d\tau \right) ^{n}  \nonumber \\
&=&{\bf I}+\sum_{n=1}^{\infty }\frac{\left( -i\right) ^{n}}{n!}
\int_{0}^{t}dt_{n}\int_{0}^{t}dt_{n-1}...\int_{0}^{t}dt_{1}{\rm T}\left\{ 
{\bf H}_{I}(t_{1})\cdots {\bf H}_{I}(t_{n})\right\}   \nonumber \\
&\equiv &{\bf I}+\sum_{n=1}^{\infty }\frac{\left( -i\right) ^{n}}{n!}{\bf U}
_{n}(t).  \label{eq:U}
\end{eqnarray}
The Dyson time-ordered product is defined with respect to any set of
operators ${\bf O}_{i}(t_{i})$ as \cite{March:book} 
\[
{\rm T}\left\{ {\bf O}_{1}(t_{1})\cdots {\bf O}_{n}(t_{n})\right\} ={\bf O}
_{\tau _{1}}(t_{\tau _{1}})\cdots {\bf O}_{\tau _{n}}(t_{\tau _{n}})\qquad
\left( t_{\tau _{1}}>t_{\tau _{2}}>...>t_{\tau _{n}}\right) .
\]
Using Eq.~(\ref{eq:H_I}) we have for the terms in the above sum: 
\[
\prod_{j=1}^{n}{\bf H}_{I}(t_{j})=\sum_{{\bf i}=1}^{K}\sum_{{
\mbox{\boldmath
$\alpha$}}=x,y,z}\sum_{{\bf k}}\bigotimes_{j=1}^{n}\sigma _{i_{j}}^{\alpha
_{j}}(t_{j})\bigotimes_{j=1}^{n}{\bf B}_{i_{j}k_{j}}^{\alpha _{j}}(t_{j}),
\]
where ${\bf i}=\{i_{1},i_{2},...,i_{n}\}$, ${\mbox{\boldmath $\alpha$}}
=\{\alpha _{1},\alpha _{2},...,\alpha _{n}\}$, and ${\bf k}
=\{k_{1},k_{2},...,k_{n}\}$. The important point to notice in this
complicated expression is that after taking the bath matrix elements $
\langle \mu |\cdots |\nu \rangle $ [because of Eq.~(\ref{eq:Amunu})], one is
left with all possible tensor products $\bigotimes_{j=1}^{n}\sigma
_{i_{j}}^{\alpha _{j}}(t_{j})$ over $n$ out of $K$ qubits. The integration
and time-ordering operation will not change this conclusion. Thus, using,
the expansion of $\sigma _{i}^{\alpha }(t)$ in Eq.~(\ref{eq:sigma_I}), after
a time $O(t^{K})$ one finds the tensor product $\bigotimes_{j=1}^{K}\sigma
_{i_{j}}^{\alpha _{j}}$, i.e., {\em all} qubits are involved (here $\alpha
_{j}=0$, corresponding to the identity matrix, is allowed). At this point
the entire {\em Pauli group} $P_{K}$ appears (all possible $4^{K+1}$ tensor
products of the $3$ Pauli matrices and the identity matrix, and the four
roots of unity \{$\pm ,\pm i$\} -- see Appendix \ref{app1}), and one has
``complete decoherence'', i.e., multiple-qubit errors over the entire system
Hilbert space. In the usual approach to QECC one does not consider such high
orders in time since one assumes that error correction can be done quickly
enough. Instead the error-analysis is usually confined to time evolution to $
O(t)$ only, which leads to ``independent decoherence'', i.e., single-qubit
errors affecting only one qubit at a time.\footnote{
In fact {\em spatially} correlated errors can also be dealt with by QECCs 
\cite{Duan:99}.} It is possible to use multiple-error-correcting quantum
codes for $O(t^{n})$ with arbitrary $n$, but these codes are rather unwieldy
(i.e., the number of encoding qubits becomes large). In the case of ``burst
errors''\ (a spatially contiguous cluster of errors such as $I\cdots
IX\cdots XI\cdots I$) some particularly efficient codes are known \cite
{Bandyopadhyay:98}.

On the other hand, a DFS that exists by virtue of a spatially symmetric
system-bath coupling, is not affected by this proliferation of errors, which
all occur in the subspace orthogonal to the DFS \cite{Lidar:PRL99}. The
assumption of spatial symmetry manifests itself in restrictions on the
coefficients $g_{ik}^{\alpha }$ appearing in the interaction Hamiltonian
[Eq. $~$(\ref{eq:H_I1})]. For example, as mentioned above, collective
decoherence corresponds to the condition $g_{ik}^{\alpha }=g_{k}^{\alpha }$ $
\forall i$, i.e., the bath cannot distinguish between the qubits \cite
{Zanardi:97c}. In this paper no such spatial symmetry assumptions will be
made. Instead, only {\em multiple-qubit} errors will be allowed to lowest
order instead of single-qubit errors. This condition will be defined more
precisely in the next section.

As for the Kraus operators, it can be seen from the calculations above that
they may be expanded as sums over tensor products of the Pauli matrices: 
\begin{eqnarray}
{\bf A}_{d}(t)=\sum_{n=1}^{4^{K+1}}a_{d,n}(t){\bf p}_{n}  \label{eq:Ai}
\end{eqnarray}
where ${\bf p}_{n}\in P_{K}$. The Kraus operators thus belong to the {\em 
group algebra} (the space of linear combinations of group elements) of $
P_{K} $ \cite{Schensted}. As alluded to in the beginning of this section,
that this expansion is possible actually follows simply from the fact that
the Pauli group forms a complete orthogonal set (with respect to the trace
inner product) for the expansion (with complex coefficients) of arbitrary $
2^{K}\times 2^{K}$ matrices. However, here we have seen how the expansion in
terms of the Pauli group (rather than some other basis) is physically
motivated by virtue of the structure of the Hamiltonian.

A simple example will now serve to illustrate the point made above about
multiple-qubit errors. Consider an interaction Hamiltonian of the form ${\bf 
H}_{I}=\sum_{i=1}^{2}\sigma _{i}^{z}\otimes B_{i}$ (on two qubits). Some
algebra suffices to show that then ${\bf A}_{d}(t)=c_{0}(t){\bf I}
_{S}+c_{1}(t)\sigma _{1}^{z}+c_{2}(t)\sigma _{2}^{z}+c_{12}(t)\sigma
_{1}^{z}\otimes \sigma _{2}^{z}$. In this case the single-qubit errors $
\sigma _{1}^{z},\sigma _{2}^{z}$ appear, as well as the multiple-qubit error 
$\sigma _{1}^{z}\otimes \sigma _{2}^{z}$. This situation does not allow for
the appearance of DFSs (unless spatial symmetry is present). Alternatively,
consider the interaction Hamiltonian ${\bf H}_{I}=(\sigma _{1}^{z}\otimes
\sigma _{2}^{z})\otimes B_{12}+(\sigma _{3}^{z}\otimes \sigma
_{4}^{z})\otimes B_{34}$ (on four qubits). In this case one finds ${\bf A}
_{d}(t)=c_{0}(t){\bf I}_{S}+c_{12}(t)\sigma _{1}^{z}\otimes \sigma
_{2}^{z}+c_{34}(t)\sigma _{3}^{z}\otimes \sigma _{4}^{z}+c_{1234}(t)\sigma
_{1}^{z}\otimes \sigma _{2}^{z}\otimes \sigma _{3}^{z}\otimes \sigma
_{4}^{z} $. Thus only multiple-qubit terms appear, and as will be shown
below, this allows for the existence of non-trivial DFSs, even though no
spatial symmetry assumptions were made.

An important example of this correlated type of system-bath interaction is
the dipolar-coupling Hamiltonian, relevant, e.g., to decoherence resulting
from spin-rotation coupling in NMR \cite{Slichter:96}.\footnote{
We thank Prof. Dieter Suter for suggesting this example.} The dipolar
Hamiltonian for a system of spins interacting with a bath of rotations is: 
\begin{equation}
H_{I}=\sum_{j,k}\frac{\gamma _{j}\gamma _{k}}{r_{jk}^{3}}\left[ {{
\mbox{\boldmath $\sigma$}}}_{j}\cdot {{\mbox{\boldmath $\sigma$}}}
_{k}-3\left( {{\mbox{\boldmath $\sigma$}}}_{j}\cdot {\bf r}_{jk}\right)
\left( {{\mbox{\boldmath $\sigma$}}}_{k}\cdot {\bf r}_{jk}\right) \right] ,
\end{equation}
where $\gamma _{j}$ is the gyromagnetic ration of spin $j$, $r_{jk}$ is the
distance beween spins $j$ and $k$, and ${\mbox{\boldmath $\sigma$}}$ is the
vector of Pauli matrices. Introducing an anistropy tensor $g_{jk}^{\alpha
\beta }$, this can be rewritten as: 
\begin{equation}
H_{I}=\sum_{j,k}\frac{\gamma _{j}\gamma _{k}}{r_{jk}^{3}}\sum_{\alpha ,\beta
=-1}^{1}g_{jk}^{\alpha \beta }\left( \sigma _{j}^{\alpha }\otimes \sigma
_{k}^{\beta }\right) Y_{2}^{-\alpha -\beta },  \label{eq:dipolar}
\end{equation}
where $Y_{l}^{m}$ are the spherical harmonics, and $\sigma ^{0}\equiv \sigma
^{z}$. Eventhough only multiple-qubit terms appear here it is necessary to
further impose anisotropy in order to obtain an example with a non-trivial
DFS, as we discuss in more detail in Sec.~\ref{Q2Z}. This is the case, e.g.,
when only $\sigma _{j}^{z}\otimes \sigma _{k}^{z}$ terms remain (i.e., $
g_{jk}^{\alpha \beta }=\delta _{\alpha 0}\delta _{\beta 0}g_{jk}$), coupled
to $Y_{2}^{0}$ rotations.

With these observations, we are now ready to study the question of DFSs in
open systems without spatial symmetry in the system-bath couplings.

\section{Decoherence-Free Subspaces from Subgroups of the Pauli Group}

\label{DFS}

We begin this section by recalling the condition for DFSs within the
framework of the Kraus operator-sum representation, derived in Ref.~\cite
{Lidar:PRL99}. We then analyze the conditions for the appearance of DFSs
when the errors are spanned by a subgroup of the Pauli group. The result is
summarized by a theorem presented at the end of the section.

\subsection{Condition for Decoherence-Free Subspaces}

A DFS is a subspace $\tilde{{\cal H}}={\rm Span}\{|\tilde{j}\rangle \}$ of
the full system Hilbert space ${\cal H}_{K}$ over which the evolution of the
density matrix is unitary. Necessary and sufficient conditions for a DFS
were derived in the Markovian case in Ref.~\cite{Lidar:PRL98} and in the
exact (non-Markovian) case in Ref.~\cite{Zanardi:97a}. A formulation of the
exact DFS condition was given in terms of the operator sum representation in
Ref.~\cite{Lidar:PRL99}, and will be briefly reviewed.

Let $\{|\tilde{j}\rangle \}$ be a set of system states satisfying: 
\begin{equation}
{\bf A}_{d}|\tilde{j}\rangle =c_{d}{\tilde{{\bf U}}}|\tilde{j}\rangle \qquad
\forall d,  \label{eq:DFS-cond}
\end{equation}
where $\tilde{{\bf U}}$ is an arbitrary, $d$-independent but possibly
time-dependent unitary transformation, and $c_{d}$ a complex constant. Under
this condition, an initially pure state belonging to ${\rm Span}[\{|\tilde{j}
\rangle \}]$, 
\[
|\psi _{{\rm in}}\rangle =\sum_{j}\gamma _{j}|\tilde{j}\rangle ,
\]
will be {\em decoherence-free}, since: 
\[
|\phi _{d}\rangle ={\bf A}_{d}|\psi _{{\rm in}}\rangle =\sum_{j}\gamma
_{j}c_{d}{\tilde{{\bf U}}}|\tilde{j}\rangle =c_{d}{\tilde{{\bf U}}}|\psi _{
{\rm in}}\rangle 
\]
so 
\[
\rho _{{\rm out}}=\sum_{d}{\bf A}_{d}\tilde{\rho}_{{\rm in}}{\bf A}
_{d}^{\dagger }=\sum_{d}c_{d}{\tilde{{\bf U}}}|\psi _{{\rm in}}\rangle
\langle \psi _{{\rm in}}|{\tilde{{\bf U}}}^{\dagger }c_{d}^{\ast }={\tilde{
{\bf U}}}|\psi _{{\rm in}}\rangle \langle \psi _{{\rm in}}|{\tilde{{\bf U}}}
^{\dagger },
\]
where we used the normalization of the Kraus operators [Eq. (\ref{eq:OSRnorm}
)] to set $\sum_{d}|c_{d}|^{2}=1$. This means that the time-evolved state $
\rho _{{\rm out}}$ is pure, and its evolution is governed by ${\tilde{{\bf U}
}}$. This argument is easily generalized to an initial mixed state $\tilde{
\rho}_{{\rm in}}=\sum_{jj^{\prime }}\rho _{jj^{\prime }}|\tilde{j}\rangle
\langle \tilde{j}^{\prime }|$, in which case $\rho _{{\rm out}}={\tilde{{\bf 
U}}}\tilde{\rho}_{{\rm in}}{\tilde{{\bf U}}}^{\dagger }$. The unitary
transformation ${\ \tilde{{\bf U}}}$ is a ``gauge freedom'' which can be
exploited in choosing a driving system Hamiltonian which implements a useful
evolution on the DFS. In the interaction picture used in the previous
section, ${\tilde{{\bf U}}}$ can be made to disappear by redefining all
Kraus operators as ${\tilde{{\bf U}}}^{\dagger }{\bf A}_{d}$. The
calculation above shows that Eq.~(\ref{eq:DFS-cond}) is a sufficient
condition for a DFS. It follows from the results of Refs.~\cite
{Zanardi:97a,Nielsen:97} that it is also a necessary condition for a DFS
(under ``generic'' conditions -- to be explained below).

Eq.~(\ref{eq:DFS-cond}), however, does not seem to be a very useful
characterization of a DFS if one does not know the explicit form of the
Kraus operators (in general this cannot be found in closed analytical form,
although they can be determined experimentally \cite{Chuang:97c}). When the
Kraus operators derive from a Hamiltonian, as in Eq.~(\ref{eq:Amunu}), an
equivalent DFS condition is \cite{Lidar:PRL99}: 
\begin{eqnarray}
{\bf S}_{\alpha }|\tilde{j}\rangle =a_{\alpha }|\tilde{j}\rangle \qquad
\forall \alpha ,
\end{eqnarray}
where the system-bath interaction Hamiltonian is written as ${\bf H}
_{I}=\sum_{\alpha }{\bf S}_{\alpha }\otimes {\bf B}_{\alpha }$ [compare to
Eq.~(\ref{eq:H_I})], with $\{{\bf S}_{\alpha }\}$ being the system
operators. To make use of this last DFS condition, one needs to introduce
assumptions about the structure of system-bath coupling, and this is how one
is led to spatial symmetry considerations \cite{Lidar:PRL98}. Here, however,
the DFS condition of Eq.~(\ref{eq:DFS-cond}) will be considered directly,
based purely on the expansion of the Kraus operators in terms of the Pauli
group elements, and without resorting to an explicit form for these
operators.

\subsection{Representation Theory Construction of Decoherence-Free States}

\label{reptheory}

When the Kraus operators are viewed as operators in the algebra of the Pauli
group, the DFS condition [Eq.~(\ref{eq:DFS-cond})] has a natural
interpretation: the decoherence free states $\{|\tilde{j}\rangle \}$ belong
to the one-dimensional irreps of the Pauli group. Motivated by this
observation we now consider a group representation theory construction of
decoherence-free states.

The general criterion for the reducibility of a representation $\{\Gamma
(G_{n})\}_{n=1}^{N}$of a finite group ${\cal G}=\{G_{n}\}$ of order $N$ is 
\cite{Cornwell:97}: 
\begin{eqnarray}
\sum_{n=1}^{N}|\chi \lbrack \Gamma (G_{n}) \rbrack |^{2}>N,
\end{eqnarray}
where $\chi $ is the character of the representation $\Gamma $ [trace of the
matrix $\Gamma (G_{n})$]. If equality holds, then the representation is
irreducible.

The full Pauli group $P_{K}$ is irreducible over the Hilbert space ${\cal H}
_{K}$ of $K$ qubits: since all Pauli matrices are traceless, only the four
elements proportional to the identity matrix contribute (see also Appendix 
\ref{app1}): 
\[
\sum_{n=1}^{4^{K+1}}|\chi \lbrack {\bf p}
_{n}]|^{2}=|2^{K}|^{2}+|-2^{K}|^{2}+|i2^{K}|^{2}+|-i2^{K}|^{2}=4^{K+1},
\]
which is just the order of $P_{K}$ (generally the direct product
representation of irreps of any direct product group is itself an irrep of
that group \cite{Cornwell:97}).

Now we come to the central assumption setting the stage for the DFSs
considered in this paper:\ {\em what if the Kraus operators belong to the
group algebra of a {\bf subgroup} }$Q${\em \ of }$P_{K}$? The motivation for
this situation could be: (i) The case in which either only higher order
errors occur, such that first-order terms of the form $I\otimes \cdots
\otimes I\otimes \sigma _{i}^{\alpha }\otimes I\otimes \cdots \otimes I$ are
absent in the Pauli-group expansion of the Kraus operators, or (ii) only
errors of {\em one} kind, either $\sigma ^{x}$, $\sigma ^{y}$, or $\sigma
^{z}$ take place. Case (i) would imply that either: (a) There are certain
cancellations involving bath matrix element terms such that first-order
system operators are absent in the expansion of Eq.~(\ref{eq:U}). This would
be a rather non-generic situation, involving a very special ``friendly''
bath; or (b) The system-bath Hamiltonian is in fact not of the form in Eq.~(
\ref{eq:H_I}), but rather involves only second order terms such as $\sigma
_{i}^{\alpha }\otimes \sigma _{j}^{\beta }$ (identity on all the rest).
\footnote{
Note that in this case the expansion of the Kraus operators in terms of
tensor products of Pauli matrices, Eq.~(\ref{eq:Ai}) remains valid.} Case
(ii) is applicable in, e.g., the case of pure phase damping (relevant to NMR 
\cite{Slichter:96}) and optical lattices using cold controlled collisions 
\cite{Jaksch:99}), where $\sigma ^{z}$ errors are dominant.

In the subgroup case under consideration, we may find non-trivial
irreducible representations (irreps) of $Q$ over ${\cal H}_{K}$ (a so-called
``subduced'' representation \cite{Schensted}). This situation can be
interesting especially if there exist 1-dimensional irreps, as known from
the general theory of DFSs \cite{Zanardi:97a,Lidar:PRL98}. As will be shown
next, the recipe for finding these DFSs uses the standard projection
operators from elementary group representation theory. The projection is
onto the subspace transforming according to a particular irrep.

First, recall the multiplicity formula for unitary irreps (which we can
always assume in this case since the Pauli group is finite): 
\begin{eqnarray}
m_{k}=\frac{1}{N}\sum_{n=1}^{N}\chi \left[ \Gamma ^{k}(G_{n})\right] ^{\ast
}\chi \left[ \Gamma (G_{n})\right] ,  \label{eq:mk}
\end{eqnarray}
where $m_{k}$ is the number of times irrep $\Gamma ^{k}$ appears in the
given reducible representation; $\chi \left[ \Gamma ^{k}(G_{n})\right] $ is
the character of the $\Gamma ^{k}$ irrep on the group element $G_{n}$; and $
\chi \left[ \Gamma (G_{n})\right] $ is the character of $G_{n}$ in the given
reducible representation $\Gamma $.

We denote a set of (orthonormal) basis-states transforming according to an
irrep $\Gamma^k$ by $\{|\psi_1^k \rangle, \ldots ,|\psi_{d_k}^k \rangle \}$.
These states span the invariant subspace of the irrep $\Gamma^k$ and
transform according to:

\begin{eqnarray}
G_{n}|\psi _{\mu }^{k}\rangle =\sum_{\nu =1}^{d_{k}}\Gamma ^{k}(G_{n})_{\nu
\mu }|\psi _{\nu }^{k}\rangle .
\label{eq:trafo}
\end{eqnarray}
Furthermore they obey the orthogonality relation: 
\begin{eqnarray}
\langle \psi _{\mu }^{l}|\psi _{\nu }^{k}\rangle =\delta _{lk}\delta _{\mu
\nu }
\label{eq:ortho}
\end{eqnarray}
Next, a projection operator onto the subspace belonging to the $d_{k}$
-dimensional irrep $k$ is given by the appropriate sum over group elements 
\cite{Cornwell:97}: 
\begin{eqnarray}
P_{\mu \nu }^{k}=\frac{d_{k}}{N}\sum_{n=1}^{N}\Gamma ^{k}(G_{n})_{\mu \nu
}^{\ast }G_{n}\;;\qquad \mu ,\nu =1,...,d_{k}
\label{eq:proj}
\end{eqnarray}
and has the following properties:

\begin{eqnarray}
P_{\mu \nu }^{k}P_{\kappa \lambda }^{l} &=&\delta _{kl}\delta _{\nu \kappa
}P_{\mu \lambda }^{k}  \nonumber \\
P_{\mu \nu }^{l}|\psi _{\lambda }^{k}\rangle  &=&\delta _{kl}\delta _{\nu
\lambda }|\psi _{\mu }^{k}\rangle {.}
\end{eqnarray}
To obtain a set of (orthonormal) basis states $\{|\psi
_{1}^{k}\rangle,\ldots ,|\psi _{d_{k}}^{k}\rangle \}$ transforming as
a set of partners in 
the basis for $\Gamma ^{k}$ from an arbitrary state $|\phi \rangle $, one
can apply the set of operators $\{P_{\mu \nu }^{k}\}$ for a fixed $\nu $
(such that $P_{\nu \nu }^{k}|\phi \rangle \neq 0$) and renormalize the
states thus obtained. Every state $|\phi \rangle $ can be expanded in terms
of basis states for the constituting irreps $\Gamma ^{k}$ as

\begin{eqnarray}
|\phi \rangle =\sum_{k}\sum_{\nu =1}^{d_{k}}\theta _{\nu }^{k}|\psi _{\nu
}^{k}\rangle ,
\label{eq:expand}
\end{eqnarray}
where $P_{\nu \nu }^{k}|\phi \rangle =\theta _{\nu }^{k}|\psi _{\nu
}^{k}\rangle $ and the summation over $k$ is over inequivalent irreps \cite
{Cornwell:97}.

Let us now consider the effect of applying the\ operators ${\bf A}
_{d}=\sum_{n}a_{d,n}G_{n}$ from the group algebra to an arbitrary state $
|\phi \rangle $:

\begin{eqnarray}
{\bf A}_{d}|\phi \rangle =\sum_{k}\sum_{\mu =1}^{d_{k}}\theta _{\mu }^{k} 
{\bf A}_{d}|\psi _{\mu }^{k}\rangle =\sum_{n=1}^{N}a_{d,n}\sum_{k}\sum_{\mu
=1}^{d_{k}}\theta _{\mu }^{k}\sum_{\nu =1}^{d_{k}} \Gamma ^{k}(G_{n}) _{\nu
\mu}|\psi _{\nu}^{k}\rangle .  \label{eq:Ai-proj}
\end{eqnarray}
We would like to find the conditions such that this transforms into the DFS
condition, Eq.~(\ref{eq:DFS-cond}). Consider the case when $\Gamma ^{k}$ are
all {\em 1-dimensional irreps}, possibly appearing with multiplicity $m_{k}$:

\begin{eqnarray}
\Gamma ^{k}(G_{n})_{\mu \nu }=\gamma _{n}^{k} \qquad \mu ,\nu =1.
\label{eq:diag}
\end{eqnarray}
In this case the indices $\mu ,\nu $ are irrelevant and we will omit them.
Then: 
\begin{eqnarray}
{\bf A}_{d}|\phi \rangle =\sum_{n=1}^{N}a_{d,n}\sum_{k}\gamma _{n}^{k}\theta
^{k}|\psi ^{k}\rangle .
\end{eqnarray}
For $|\phi \rangle $ to be a decoherence free state, one would like to have
this proportional to $|\phi \rangle =\sum_{k}\theta ^{k}|\psi ^{k}\rangle $
[as in the original expansion of Eq.~(\ref{eq:expand})]. However, this does
not work because of the presence of $\gamma _{n}^{k}$ in the sum. We thus
see that the initial function $|\phi \rangle $ must be restricted to be one
of the basis-states $|\psi ^{k}\rangle $. Then, with

\begin{eqnarray}
c_{d}^{k}\equiv \sum_{n=1}^{N}a_{d,n}\gamma _{n}^{k}.  \label{eq:c_d}
\end{eqnarray}
we have finally:

\begin{eqnarray}
{\bf A}_{d}|\psi ^{k}\rangle =c_{d}^{k}|\psi ^{k}\rangle .  \label{eq:phiDFS}
\end{eqnarray}
At this point it is useful to introduce another index $z$ for the
multiplicity of the irrep $k$, i.e. $z=1\ldots m_{k}$. The Hilbert space of $
K$-qubit states splits into invariant one-dimensional subspaces $V_{z}^{k}$
that are spanned by (fixed) basis states $|\psi _{z}^{k}\rangle $. Each of
the $|\psi ^{k}\rangle $ in Eq.~(\ref{eq:expand}) is a linear combination of
the $|\psi _{z}^{k}\rangle $:

\begin{eqnarray}
|\psi ^{k}\rangle =\sum_{z=1}^{m_{k}}\theta _{z}^{k}|\psi _{z}^{k}\rangle ,
\label{eq:sameirrep}
\end{eqnarray}
[Because of Eq.~(\ref{eq:expand}), the $\theta _{z}^{k}$ depend on the
initial state $|\phi \rangle $.] Thus for $|\phi \rangle $ to be a
decoherence-free state, it is allowed to be an arbitrary superposition
inside copies of a {\em given} irrep (different $z$'s), but not to be a
superposition between different irreps (different $k$'s). In particular we
have within each copy of the irrep $\Gamma ^{k}$:

\begin{eqnarray}
{\bf A}_{d}|\psi _{z}^{k}\rangle =c_{d}^{k}|\psi _{z}^{k}\rangle \;;\qquad
z=1..m_{k}.  \label{eq:cik-DFS}
\end{eqnarray}
This is just the DFS condition, Eq.~(\ref{eq:DFS-cond}) with the $\{|\psi
_{z}^{k}\rangle \}$ being the basis states for the DFS. Therefore Eq.~(\ref
{eq:diag}) is a {\em sufficient }condition for a DFS, provided that our
initial state satisfies the condition that it is a superposition of states
within a {\em fixed} irrep, Eq.~(\ref{eq:sameirrep}).

It will now be shown that Eq.~(\ref{eq:diag}) is also a {\em necessary }
condition for a DFS under the ``genericity'' assumption that the error
coefficients $\{a_{d,n}\}$ are arbitrary. In other words, it will be shown
under these conditions that, if a set of basis-states $\{|\tilde{j}\rangle \}
$ satisfies the DFS-condition Eq.~(\ref{eq:DFS-cond}) then the $\{|\tilde{j}
\rangle \}$ belong to the invariant subspace of some one-dimensional irrep
of our subgroup.

Assume that the ${\bf A}_{d}$ have been redefined to incorporate the
(constant) unitary transformation $\tilde{{\bf U}}$ such that Eq.~(\ref
{eq:DFS-cond}) becomes ${\bf A}_{d}|\tilde{j}\rangle =c_{d}|\tilde{j}\rangle 
$. Expand the state $|\tilde{j}\rangle $ as in Eq.~(\ref{eq:expand}):
\footnote{
For notational simplicity we avoid introducing another index for the
multiplicity of the irrep here. That such superpositions are allowed for DF
states is clear from Eq.~(\ref{eq:sameirrep}).} 
\begin{eqnarray}
|\tilde{j}\rangle =\sum_{k}\sum_{\mu =1}^{d_{k}}\theta _{\mu }^{\tilde{j}
,k}|\psi _{\mu }^{k}\rangle 
 \label{eq:expandj}
\end{eqnarray}
where $P_{\mu \mu }^{k}|\phi \rangle =\theta _{\mu }^{\tilde{j},k}|\psi
_{\mu }^{k}\rangle $. Now using Eq.~(\ref{eq:Ai-proj}): 
\begin{eqnarray}
{\bf A}_{d}|\tilde{j}\rangle  &=&c_{d}|\tilde{j}\rangle
=c_{d}\sum_{k}\sum_{\mu =1}^{d_{k}}\theta _{\mu }^{\tilde{j},k}|\psi _{\mu
}^{k}\rangle  \\
&=&\sum_{k}\sum_{\mu =1}^{d_{k}}\theta _{\mu }^{\tilde{j},k}{\bf A}_{d}|\psi
_{\mu }^{k}\rangle =\sum_{k}\sum_{\mu =1}^{d_{k}}\theta _{\mu }^{\tilde{j}
,k}\sum_{n}a_{d,n}G_{n}|\psi _{\mu }^{k}\rangle   \nonumber \\
&=&\sum_{k}\sum_{\mu =1}^{d_{k}}\theta _{\mu }^{\tilde{j},k}\sum_{n}a_{d,n}
\sum_{\lambda =1}^{d_{k}}\Gamma ^{k}(G_{n})_{\lambda \mu }|\psi _{\lambda
}^{k}\rangle   \label{eq:long1}
\end{eqnarray}
and taking inner products [using (\ref{eq:ortho})] 
\begin{eqnarray}
\langle \psi _{\sigma }^{l}|{\bf A}_{d}|\tilde{j}\rangle 
&=&c_{d}\sum_{k}\sum_{\mu =1}^{d_{k}}\theta _{\mu }^{\tilde{j},k}\langle
\psi _{\sigma }^{l}|\psi _{\mu }^{k}\rangle =c_{d}\theta _{\sigma }^{\tilde{j
},l}  \nonumber \\
&=&\sum_{k}\sum_{\mu =1}^{d_{k}}\theta _{\mu }^{\tilde{j},k}\sum_{n}a_{d,n}
\sum_{\lambda =1}^{d_{k}}\Gamma ^{k}(G_{n})_{\lambda \mu }\langle \psi
_{\sigma }^{l}|\psi _{\lambda }^{k}\rangle =\sum_{\mu =1}^{d_{l}}\theta
_{\mu }^{\tilde{j},l}\sum_{n}a_{d,n}\Gamma ^{l}(G_{n})_{\sigma \mu }.
\label{eq:long}
\end{eqnarray}
Using this result we would like to show that the $\Gamma ^{l}(G_{n})$ that
appear here must be one-dimensional irreps. Let us establish ``generic''
conditions for this purpose.

Eq.~(\ref{eq:long}) can be rewritten as an eigenvalue equation: 
\begin{eqnarray}
{\cal A}_{d}^{l}\overrightarrow{\theta _{j}^{l}}=c_{d}\overrightarrow{\theta
_{j}^{l}},
\label{eq:eigen}
\end{eqnarray}
where 
\begin{eqnarray}
{\cal A}_{d}^{l} &\equiv &\sum_{n}a_{d,n}\Gamma ^{l}(G_{n}) \\
\overrightarrow{\theta _{j}^{l}} &\equiv &(\theta _{1}^{\tilde{j},l},\ldots
,\theta _{d_{l^{\prime }}}^{\tilde{j},l}).
\end{eqnarray}
The vector $\overrightarrow{\theta _{j}^{l}}$ may be zero for a given irrep $
\Gamma ^{l}$, in which case Eq.~(\ref{eq:eigen}) is trivially satisfied. Let
us assume this is not the case for some $l$ [it cannot be the case for {\em 
all} $l$, by Eq.~(\ref{eq:expandj})]. Then the most general way in which
Eq.~(\ref{eq:eigen}) can be satisfied, is for $\overrightarrow{\theta
_{j}^{l}}$ to be an eigenvector of ${\cal A}_{d}^{l}$ for all codewords $|
\tilde{j}\rangle $, with eigenvalue $c_{d}$. However, while this is the most
general condition, it is {\em non-generic}. By {\em generic} we mean that we
take the errors to be {\em arbitrary}, i.e., we do not want to make any
assumptions on the $a_{d,n}$. Now, if the eigenvalue eqnarray were to be
satisfied, the vector of coefficients $\overrightarrow{\theta _{j}^{l}}$
would have to be ``special''. In other words, {\em it would have to be
adjusted to be an eigenvector of} ${\cal A}_{d}^{k}$. To make this
adjustment would require two conditions: (i) Having a priori knowledge of
the $a_{d,n}$, (ii) Being able to control $\overrightarrow{\theta _{j}^{l}}$
. We would like to avoid assuming (i) because fine-tuning the bath is
physically unacceptable. In contrast, control of $\overrightarrow{\theta
_{j}^{l}}$ is certainly desirable. However, we would like to avoid the
situation where only certain special choices of $\overrightarrow{\theta
_{j}^{l}}$, compatible with specific bath parameters, yield decoherence free
states $|\tilde{j}\rangle $.\footnote{
This statement of what are generic conditions that lead to a DFS is very
similar to that in Ref.~\cite{Lidar:PRL98}.} We thus conclude that to avoid
fine-tuning of the bath parameters and/or special initial conditions, ${\cal 
A}_{d}^{l}$ must be proportional to the identity. But since $\Gamma ^{l}$ is
an irrep this is only possible if it is one-dimensional, i.e. $\Gamma
^{l}(G_{n})_{\mu \nu }=\gamma _{n}^{l}$, $\mu ,\nu =1$ and $
c_{d}=\sum_{n}a_{d,n}\gamma _{n}^{l}$. In addition we see that $c_{d}$ can
only be $l$-independent if the DFS states $|\tilde{j}\rangle $ are spanned 
{\em only} by basis states of copies of the {\em same} irrep $\Gamma ^{l}$.
Q.E.D.

To summarize:

{\it Theorem 1}: Suppose that the Kraus operators belong to the group
algebra of some group ${\cal G}=\{G_{n}\}$, i.e., ${\bf A}
_{d}=\sum_{n=1}^{N}a_{d,n}G_{n}$. If a set of states $\{|\tilde{j}\rangle \}$
belong to a given {\em one-dimensional} irrep $\Gamma ^{k}$ of ${\cal G}$,
then the DFS condition ${\bf A}_{d}|\tilde{j}\rangle =c_{d}|\tilde{j}\rangle 
$ holds. If no assumptions are made on the bath coefficients $\{a_{d,n}\}$,
then the DFS condition ${\bf A}_{d}|\tilde{j}\rangle =c_{d}|\tilde{j}\rangle 
$ implies that $|\tilde{j}\rangle $ belongs to a {\em one-dimensional} irrep 
$\Gamma ^{k}$ of ${\cal G}$.

For completeness we give in Appendix \ref{app2} an example of the
``non-generic DFSs'', which result from ``accidentally'' satisfying Eq.~(\ref
{eq:eigen}) with irreps of dimension greater than one.

\section{Examples of Subgroups with Decoherence Free States}

\label{Examples}

The general considerations from the previous section will now be illustrated
with some examples. To simplify the notation, let $X,Y,Z$ represent the $
\sigma ^{x},\sigma ^{y},\sigma ^{z}$ Pauli matrices, and let us drop the
tensor product symbol (i.e., let $ZI\equiv Z\otimes I$, $X^{2}\equiv
X\otimes X$, etc.). Also, we will ignore normalization factors in this
section.

\subsection{Abelian Subgroups}

The simplest non-trivial example of a subgroup is found already for $K=2$
qubits: 
\begin{eqnarray}
Q_{Z}=\{I^{2},ZI,IZ,Z^{2}\}.
\end{eqnarray}
This subgroup (generated by $ZI$ and $IZ$) describes phase damping.

As another simple example, let $K=4$ qubits and consider the following
subgroup: 
\[
Q_{X}=\{I^{4},X^{2}I^{2},I^{2}X^{2},X^{4}\}.
\]
Physically, this would correspond to the error process where bit flips
happen on certain clusters of two or four qubits only (note that $XIXI$ and $
IXIX$ were left out -- this case will be considered in the sequel paper \cite
{Lidar:00b}).

Another example is 
\[
Q_{4}=\{I^{4},X^{4},Y^{4},Z^{4}\},
\]
with all Pauli errors occurring just on clusters of 4 qubits. $Q_{Z}$, $Q_{X}
$ and $Q_{4}$ are isomorphic and Abelian. All elements of these subgroups,
except $I^{4}$, are traceless. $I^{4}$ has trace 16, so that $
\sum_{n=1}^{4}|\chi \lbrack \Gamma (G_{n})]|^{2}=256>4$ and thus the natural
representation of these subgroups on $4$ qubits is reducible. Since they are
Abelian, they have only 1-dimensional irreps. These irreps are given in the
following table, expressed in terms of the elements of $Q_{X}$: 
\begin{eqnarray}
\begin{tabular}{|r||r|r|r|r|}
\hline
& $I^{4}$ & $X^{2}I^{2}$ & $I^{2}X^{2}$ & $X^{4}$ \\ \hline\hline
$\Gamma ^{1}$ & 1 & 1 & 1 & 1 \\ \hline
$\Gamma ^{2}$ & 1 & 1 & -1 & -1 \\ \hline
$\Gamma ^{3}$ & 1 & -1 & 1 & -1 \\ \hline
$\Gamma ^{4}$ & 1 & -1 & -1 & 1 \\ \hline
\end{tabular}
\label{eq:mult-table}
\end{eqnarray}
Motivated by Theorem 1, this reducibility implies the existence of DFSs, as
long as the Kraus operators belong to the group algebra of these subgroups.

\subsubsection{The Subgroup $Q_{X}$}

Consider the case of $Q_{X}$, i.e., assume that the Kraus operators can be
written as 
\begin{eqnarray}
{\bf A}_{d}=a_{d,0}I^{4}+a_{d,1}X^{2}I^{2}+a_{d,2}I^{2}X^{2}+a_{d,3}X^{4}
\label{eq:generalA}
\end{eqnarray}
[the coefficients $a_{d,j}$ are of course constrained by the normalization
condition Eq.~(\ref{eq:OSRnorm})].

Using the general arguments of Sec.~\ref{reptheory} and in particular Eq.~(
\ref{eq:proj}), we can just read off the matrix elements of the 4
(1-dimensional) irreps from the table in Eq.~(\ref{eq:mult-table}). Thus the
4 projection operators are: 
\begin{eqnarray}
P^{1} &=&I^{4}+X^{2}I^{2}+I^{2}X^{2}+X^{4}\qquad
P^{2}=I^{4}+X^{2}I^{2}-I^{2}X^{2}-X^{4}  \nonumber \\
P^{3} &=&I^{4}-X^{2}I^{2}+I^{2}X^{2}-X^{4}\qquad
P^{4}=I^{4}-X^{2}I^{2}-I^{2}X^{2}+X^{4}
\end{eqnarray}
The multiplicity of each of the four 1-dimensional irreps in the reducible
representation generated here by the $K=4$ qubits, is 4. To see this, recall
the multiplicity formula Eq.~(\ref{eq:mk}). In the present case, the given
representation yields $\chi =\{16,0,0,0\}$ (for $
I^{4},X^{2}I^{2},I^{2}X^{2},X^{4}$ respectively) and so, with $\chi
^{k}(I^{4})=1$, $m_{k}=\frac{1}{4}\chi ^{k}(I^{4})16=4$ for all $k$.

Now, let us explicitly find the decoherence free states. To do so we can
pick an arbitrary, convenient 4-qubit state and project it onto a given
irrep. For example, starting with $|0000\rangle $: 
\begin{eqnarray}
P^{1}|0000\rangle &=&|0000\rangle +|1100\rangle +|0011\rangle +|1111\rangle
\equiv |\psi _{1}^{1}\rangle  \nonumber \\
P^{2}|0000\rangle &=&|0000\rangle +|1100\rangle -|0011\rangle -|1111\rangle
\equiv |\psi _{1}^{2}\rangle  \nonumber \\
P^{3}|0000\rangle &=&|0000\rangle -|1100\rangle +|0011\rangle -|1111\rangle
\equiv |\psi _{1}^{3}\rangle  \nonumber \\
P^{4}|0000\rangle &=&|0000\rangle -|1100\rangle -|0011\rangle +|1111\rangle
\equiv |\psi _{1}^{4}\rangle .
\end{eqnarray}
Each of these 4 states belongs to a different irrep, and thus to a different
DFS, which can be verified by applying an arbitrary Kraus operator, as in
Eq.~(\ref{eq:generalA}). E.g., 
\begin{eqnarray}
{\bf A}_{d}|\psi _{1}^{1}\rangle &=&a_{d,0}\left( |0000\rangle +|1100\rangle
+|0011\rangle +|1111\rangle \right)  \nonumber \\
&&+a_{d,1}(|1100\rangle +|0000\rangle +|1111\rangle +|0011\rangle ) 
\nonumber \\
&&+a_{d,2}(|0011\rangle +|1111\rangle +|0000\rangle +|1100\rangle ) 
\nonumber \\
&&+a_{d,3}(|1111\rangle +|0011\rangle +|1100\rangle +|0000\rangle ) 
\nonumber \\
&=&\left( a_{d,0}+a_{d,1}+a_{d,2}+a_{d,3}\right) |\psi _{1}^{1}\rangle .
\end{eqnarray}
Similarly: 
\begin{eqnarray}
{\bf A}_{d}|\psi _{1}^{2}\rangle &=&\left(
a_{d,0}+a_{d,1}-a_{d,2}-a_{d,3}\right) |\psi _{1}^{2}\rangle ,  \nonumber \\
{\bf A}_{d}|\psi _{1}^{3}\rangle &=&\left(
a_{d,0}-a_{d,1}+a_{d,2}-a_{d,3}\right) |\psi _{1}^{3}\rangle ,  \nonumber \\
{\bf A}_{d}|\psi _{1}^{4}\rangle &=&\left(
a_{d,0}-a_{d,1}-a_{d,2}+a_{d,3}\right) |\psi _{1}^{4}\rangle .
\end{eqnarray}
This is in agreement with Eq.~(\ref{eq:cik-DFS}).

Now, recall that each irrep appears 4 times. This means we should be able to
find 3 more independent states belonging to each of the irreps. Indeed, by
performing projections on the states $|0001\rangle $, $|0100\rangle $, $
|1001\rangle $ (using $|0010\rangle $ and $|1000\rangle $ does not produce
new states) we obtain the complete basis for the DFSs. E.g., 
\begin{eqnarray}
P^{1}|0001\rangle  &=&|0001\rangle +|1101\rangle +|0010\rangle +|1110\rangle
\equiv |\psi _{2}^{1}\rangle ,  \nonumber \\
P^{1}|0100\rangle  &=&|0100\rangle +|1000\rangle +|0111\rangle +|1011\rangle
\equiv |\psi _{3}^{1}\rangle ,  \nonumber \\
P^{1}|1001\rangle  &=&|1001\rangle +|0101\rangle +|1010\rangle +|0110\rangle
\equiv |\psi _{4}^{1}\rangle ,
\end{eqnarray}
and again 
\begin{eqnarray}
{\bf A}_{d}|\psi _{2}^{1}\rangle  &=&\left[
a_{d,0}I^{4}+a_{d,1}X^{2}I^{2}+a_{d,2}I^{2}X^{2}+a_{d,3}X^{4}\right] \left(
|0001\rangle +|1101\rangle +|0010\rangle +|1110\rangle \right)   \nonumber \\
&=&\left( a_{d,0}+a_{d,1}+a_{d,2}+a_{d,3}\right) |\psi _{2}^{1}\rangle ,
\end{eqnarray}
with similar results for the other states. All of this is in agreement with
the general results of Sec.~\ref{reptheory}. Finally, we may consider an
arbitrary superposition of decoherence free states taken from the multiple
appearances of a given irrep, $|\phi ^{k}\rangle =\sum_{z=1}^{4}\theta
_{z}^{k}|\psi _{z}^{k}\rangle $, and this will again be decoherence-free.

\subsubsection{The Subgroup $Q_{4}$}

In this case the Kraus operators can be written as 
\begin{eqnarray}
{\bf A}_{d}=a_{d,0}I^{4}+a_{d,1}X^{4}+a_{d,2}Y^{4}+a_{d,3}Z^{4}.
\end{eqnarray}

Again, using the general arguments of Sec.~\ref{reptheory}, in the case of $
Q_{4}$ we can just read off the matrix elements of the 4 (1-dimensional)
irreps from the table in Eq.~(\ref{eq:mult-table}). Thus the 4 projection
operators are: 
\begin{eqnarray}
P^{1} &=&I^{4}+X^{4}+Y^{4}+Z^{4}\qquad P^{2}=I^{4}+X^{4}-Y^{4}-Z^{4} 
\nonumber \\
P^{3} &=&I^{4}-X^{4}+Y^{4}-Z^{4}\qquad P^{4}=I^{4}-X^{4}-Y^{4}+Z^{4}.
\end{eqnarray}
Using the multiplicity formula, Eq.~(\ref{eq:mk}), the given representation
again yields $\chi =\{16,0,0,0\}$ (for $I^{4},X^{4},Y^{4},Z^{4}$
respectively) and so once more $m_{k}=\frac{1}{4}\chi ^{k}(I^{4})16=4$ for
all $k$.

To find the decoherence free states let us start again with $|0000\rangle $.
We find: 
\begin{eqnarray}
P^{1}|0000\rangle &=& 2(|0000\rangle +|1111\rangle ) \equiv
|\psi_{1}^{1}\rangle  \nonumber \\
P^{2}|0000\rangle &=& |0000\rangle +|1111\rangle -|1111\rangle -|0000\rangle
= 0  \nonumber \\
P^{3}|0000\rangle &=&|0000\rangle -|1111\rangle +|1111\rangle -|0000\rangle
= 0  \nonumber \\
P^{4}|0000\rangle &=& 2(|0000\rangle -|1111\rangle ) \equiv |\psi
_{1}^{4}\rangle .
\end{eqnarray}
The vanishing of the projections of $P^{2}$ and $P^{3}$ implies that $
|0000\rangle $ has no components in the irreps $\Gamma ^{2}$ and $\Gamma
^{3} $. Thus a different starting state is needed, e.g., $|0001\rangle$.
Then 
\begin{eqnarray}
P^{2}|0001\rangle &=& 2(|0001\rangle +|1110\rangle ) \equiv |\psi
_{1}^{2}\rangle  \nonumber \\
P^{3}|0001\rangle &=& 2(|0001\rangle -|1110\rangle ) \equiv |\psi
_{1}^{3}\rangle.
\end{eqnarray}
That these states are decoherence free, is again easily verified by
application of an arbitrary Kraus operator, e.g.: 
\begin{eqnarray}
{\bf A}_{d}|\psi _{1}^{2}\rangle &=&\left[
a_{d,0}I^{4}+a_{d,1}X^{4}+a_{d,2}Y^{4}+a_{d,3}Z^{4}\right] 2\left(
|0001\rangle +|1110\rangle \right)  \nonumber \\
&=&\left( a_{d,0}+a_{d,1}-a_{d,2}-a_{d,3}\right) |\psi _{1}^{2}\rangle ,
\end{eqnarray}
etc. The full DFS corresponding to the projection $P^{1}$ is found by
applying $P^{1}$ to the initial states $|0011\rangle, |0101\rangle ,
|1001\rangle$: 
\begin{eqnarray}
P^{1}|0011\rangle &=& 2(|0011\rangle +|1100\rangle ) \equiv |\psi
_{2}^{1}\rangle  \nonumber \\
P^{1}|0101\rangle &=& 2(|0101\rangle +|1010\rangle ) \equiv |\psi
_{3}^{1}\rangle  \nonumber \\
P^{1}|1001\rangle &=& 2(|1001\rangle +|0110\rangle ) \equiv |\psi
_{4}^{1}\rangle,
\end{eqnarray}
in addition to $|\psi _{1}^{1}\rangle$ above.

Since the decoherence process described by $Q_{4}$ is different from that of 
$Q_{X}$, the decoherence free states are, not surprisingly, different in the
two cases.

\subsubsection{The Subgroup $Q_{Z}$}

As another example of an Abelian subgroup, assume now that the Kraus
operators, for $K=2$ qubits, can be written as 
\begin{eqnarray}
{\bf A}_{d}=a_{d,0}I^{2}+a_{d,1}ZI+a_{d,2}IZ+a_{d,3}Z^{2}.
\end{eqnarray}
The 4 projection operators are thus: 
\begin{eqnarray}
P^{1} &=&I^{2}+ZI+IZ+Z^{2}\qquad P^{2}=I^{2}+ZI-IZ-Z^{2}  \nonumber \\
P^{3} &=&I^{2}-ZI+IZ-Z^{2}\qquad P^{4}=I^{2}-ZI-IZ+Z^{2}
\end{eqnarray}
In this case, the given representation on 2 qubits yields $\chi =\{4,0,0,0\}$
(for $I^{2},ZI,IZ,Z^{2}$ respectively) and so $m_{k}=\frac{1}{4}\chi
^{k}(I^{2})4=1$ for all $k$. Thus as expected (since the representation is
4-dimensional), the multiplicity of each of the four 1-dimensional irreps is
1.

Let us again explicitly find the decoherence free states: 
\begin{eqnarray}
P^{1}|00\rangle  &=&4|00\rangle \equiv |\psi ^{1}\rangle   \nonumber \\
P^{2}|01\rangle  &=&4|01\rangle \equiv |\psi ^{2}\rangle   \nonumber \\
P^{3}|10\rangle  &=&4|10\rangle \equiv |\psi ^{3}\rangle   \nonumber \\
P^{4}|11\rangle  &=&4|11\rangle \equiv |\psi ^{4}\rangle .
\end{eqnarray}
And indeed: 
\[
{\bf A}_{d}|\psi ^{k}\rangle =\left( a_{d,0}+a_{d,1}+a_{d,2}+a_{d,3}\right)
|\psi ^{k}\rangle ,\quad k=1,...,4.
\]
This means that each of the 4 ``computational basis states'' $|\psi
^{k}\rangle $ is {\em by itself} a DFS. However, since these DFSs belong to
different irreps, a superposition is not decoherence-free. This agrees with
the well known fact that phase damping leads to decay of the off diagonal
elements of the density matrix in the computational basis, but does not
cause any population decay.

\subsubsection{The Subgroup $Q_{2Z}$}

\label{Q2Z}

As a final example of an Abelian subgroup, let us return to the anistropic
dipolar-coupling Hamiltonian [Eq.~(\ref{eq:dipolar})] discussed in Sec.~\ref
{Evolution}. Note first that it is necessary to transform from the $
\sigma^\pm$ basis used there to $\sigma^{x,y}$ in order for our Pauli
group-based discussion to apply. Having done that, it is clear that unless
anisotropy is imposed this Hamiltonian generates the entire Pauli group,
since all bilinear combinations $\sigma^\alpha \otimes \sigma^\beta$ appear
in it. Assume therefore that we have a 4-spin molecule constrained to rotate
only about the $z$-axis. This amounts to setting $g_{jk}^{\alpha \beta
}=\delta _{\alpha 0}\delta _{\beta 0}g_{jk}$ in Eq.~(\ref{eq:dipolar}), so
that only $\sigma _{j}^{z}\otimes \sigma _{k}^{z}$ terms remain. The
corresponding subgroup is 
\begin{equation}
Q_{2Z}=\{IIII,ZZII,ZIIZ,IIZZ,ZIZI,IZZI,IZIZ,ZZZZ\}.  \label{eq:Q2Z}
\end{equation}
To find the DFS under $Q_{2Z}$, construct the projector $P^1=\frac{1}{8}
\sum_{q \in Q_{2Z}}q$ corresponding to the identity irrep of $Q_{2Z}$.
Applying this projector to the initial states $|0000\rangle$ and $
|1111\rangle$ we find a 2-dimensional DFS, spanned by these two states. This
DFS thus encodes a single qubit.

\subsection{Non-Abelian Subgroups?}

It would have been interesting to find examples of non-Abelian subgroups
which have 1-dimensional irreps and thus support a DFS. However, no such
subgroups exist in the case of the Pauli group, as we now prove.

Each two elements of the Pauli group $P_{K}$ either commute or anticommute
(Appendix \ref{app1}). Let $Q$ be a non-Abelian subgroup of $P_{K}$. Then
there must be at least two elements of $Q$, say $q_{1}$ and $q_{2}$, that
anticommute. Assume that the state $|i\rangle $ belongs to a 1-dimensional
irrep $\tilde{\Gamma}$ of $Q$. Then $\tilde{\Gamma}(q_{1})|i\rangle
=c_{1}|i\rangle $ and $\tilde{\Gamma}(q_{2})|i\rangle =c_{2}|i\rangle $,
where $c_{1},c_{2}$ are numbers. Now, by assumption $\tilde{\Gamma}
(q_{2}q_{1})=\tilde{\Gamma}(-q_{1}q_{2})$. Therefore $\tilde{\Gamma}
(q_{1}q_{2})|i\rangle =\tilde{\Gamma}(q_{1})\tilde{\Gamma}(q_{2})|i\rangle
=c_{1}c_{2}|i\rangle $, and also $\tilde{\Gamma}(q_{1}q_{2})|i\rangle =
\tilde{\Gamma}(-q_{2}q_{1})|i\rangle =\tilde{\Gamma}(-q_{2})\tilde{\Gamma}
(q_{1})|i\rangle =c_{1}\tilde{\Gamma}(-q_{2})|i\rangle $. If $\tilde{\Gamma}
(-q_{2})=-\tilde{\Gamma}(q_{2})$ then we have $\tilde{\Gamma}
(q_{1}q_{2})|i\rangle =-c_{1}c_{2}|i\rangle $ so that $c_{1}c_{2}=-c_{1}c_{2}
$. This implies that at least one of $c_{1}$ and $c_{2}$ is zero. However,
this cannot be true since the representation is {\em unitary}. Is there
another possibility? Note that $\tilde{\Gamma}(-q_{2})=\tilde{\Gamma}(-{\bf I
}q_{2})=\tilde{\Gamma}(-{\bf I})\tilde{\Gamma}(q_{2})$, so the question
boils down to the value of $\alpha $ in $\tilde{\Gamma}(-{\bf I})=\alpha 
\tilde{\Gamma}({\bf I})$. But since $(-{\bf I})(-{\bf I})={\bf I}$ it
follows that $\tilde{\Gamma}(-{\bf I})\tilde{\Gamma}(-{\bf I})=\tilde{\Gamma}
({\bf I})=1$, so that $\tilde{\Gamma}(-{\bf I})=\pm 1$. Assume then that the
other case, $\tilde{\Gamma}(-{\bf I})=1$, holds. Let us use Eq.~(\ref{eq:mk}
) while recalling that only the four multiples of the identity have
non-vanishing trace:

\begin{eqnarray}
m_k &=& \frac{1}{N}\sum_{n=1}^{N}\chi \left[ \Gamma ^{k}({\bf p}_{n})\right]
^* \chi \left[ \Gamma ({\bf p}_{n})\right]  \nonumber \\
&=& \frac{1}{N} \left( \chi\left[\Gamma ^{k}({\bf I})\right]^* (2^K)+\chi 
\left[\Gamma ^{k}(-{\bf I})\right]^* (-2^K)+\chi\left[\Gamma ^{k}(i {\bf I}) 
\right]^* (i 2^K) +\chi\left[\Gamma ^{k}(-i {\bf I})\right]^* (-i 2^K)
\right) .  \label{eq:m_k-calc}
\end{eqnarray}
Since the irrep $\Gamma^{k}$ is 1-dimensional, $\chi \left[\Gamma^k \right]
= \Gamma^k$, i.e., the character is the element itself. Now let $\Gamma^{k}=
\tilde{\Gamma}$. Then since $\tilde{\Gamma}(-{\bf I}) = 1$, and using $
\Gamma ^{k}(-i {\bf I}) = \tilde{\Gamma}(-{\bf I})\tilde{\Gamma}(i{\bf I})$,
we find $\tilde{m}=0$. Therefore such irreps do not appear at all.

Thus an anti-commuting pair of elements in $Q$ is incompatible with a
1-dimensional irrep, so that if $Q$ has a 1-dimensional irrep, it must be
Abelian.\footnote{
We thank Dr. P. Zanardi for discussions regarding this point.}

Recall that the DFS condition of theorem 1 applies to arbitrary groups.
Groups other than the Pauli group may support non-Abelian subgroups with
1-dimensional irreps (the above proof relied strongly on a property specific
to the Pauli group, that its elements either commute or anticommute).
However, at least within the Hamiltonian framework expounded in Secs.~\ref
{Hamiltonians} and \ref{Evolution}, it is the Pauli group which appears
naturally for the group algebra to which the Kraus operators belong.

\section{Dimension of the Decoherence-Free Subspaces}

\label{dimension}

We showed in the previous section that for the Pauli group, DFSs can exist
only for Abelian subgroups. This observation allows us to calculate the
dimension of these DFSs. Recall from the general discussion in Sec.~\ref
{reptheory} that in the generic case a superposition of states belonging to
different irreps will decohere, whereas a superposition of states within
copies of a given irrep will be decoherence-free (see also the examples in
the previous section). Also, by the Abelian property, each such copy only
supports a single decoherence free state. Hence {\em the dimension of the
DFS associated with a given irrep $\Gamma ^{k}$ is simply its multiplicity $
m_{k}$}.

Let $Q$ be an order-$N$ Abelian subgroup of the Pauli group on $K$ qubits.
Using Eq.~(\ref{eq:m_k-calc}) and ${\Gamma^k}(-{\bf I}) = \pm 1$ again, we
have two (and only two) cases: (i) If $\Gamma ^{k}(-{\bf I}) = 1$ then $m_k=0
$, so such irreps do not support a DFS. (ii) If $\Gamma ^{k}(- {\bf I}) = -1$
then

\[
m_{k}=2^{K+2}/N. 
\]
This shows that all irreps that support a DFS have the same multiplicity,
and thus all these DFSs have the same dimension.

If the subgroup does not include elements with the $\pm 1,\pm i$ factors, as
in the examples in Sec.~\ref{Examples}, then only the term $\Gamma ^{k}({\bf 
I})$ appears in Eq.~(\ref{eq:m_k-calc}), and consequently

\[
m_{k}=2^{K}/N\qquad {\rm no}\;\{\pm 1,\pm i\}\;{\rm {factors}.}
\]
In any case, the dimension of the DFS is inversely proportional to the order
of the subgroup. This implies a trade-off between the number of errors that
can be dealt with by the code ($N$) and the number of decoherence free
qubits ($\log _{2}m_{k}$).

As an interesting corollary we see that largest Abelian subgroup of the
Pauli group has order $2^{K+2}$ (since $m_{k}\geq 1$ implies $N\leq 2^{K+2}$
). Examples of such subgroups are:

\begin{itemize}
\item  {The group generated by all the single qubit $X$'s (or $Y$'s or $Z$
's), with $\pm 1,\pm i$.}

\item  {The group generated by $XXII...II$, $YYII...II$, $ZZII...II$, $
IIXXII...II$, $IIYYII...II$, $IIZZII...II$, ..., $II...IIZZ$, with $\pm
1,\pm i$.}
\end{itemize}

These groups support only 1-dimensional DFSs. The last group is relevant for
errors due to exchange on pairs of identical qubits \cite{Ruskai:99}, and we
see that the corresponding decoherence free state is automatically immune to
exchange errors. (See Ref.~\cite{Lidar:PRA99Exchange} for a discussion of
protection of DFSs against exchange errors arising in the spatially
correlated collective decoherence case.)

\section{Summary and Conclusions}

\label{Summary}

Decoherence-free subspaces (DFSs) are associated most commonly with the
existence of a spatial symmetry in the system-bath coupling, as in the
collective decoherence model. Here we have considered the case when no such
symmetry is assumed, and have shown that one can nevertheless find DFSs
under certain conditions. The essential assumptions are that either to
lowest order only {\em multiple}-qubit errors are possible, meaning that the
bath can only couple to multiple system excitations; or, that only one type
of error process (such as phase-damping) occurs, which can be relevant for
the NMR quantum computer schemes and optical lattices (or any other
realization where scattering-induced phase-shifts are the dominant
decoherence mechanism). In either case, instead of the full Pauli group of
errors, only a subgroup needs to be considered. Barring certain non-generic
cases, the DFSs then correspond to states that transform according to the $1$
-dimensional irreducible representations of such a subgroup. This
characterization of DFSs, while formally similar to previous results, is
different in that it trades the assumption of spatial symmetry for one of
multiple-qubit coupling to the bath.

We show in a sequel paper \cite{Lidar:00b} how to perform universal fault
tolerant quantum computation on the DFSs found in this paper using only one-
and two-body Hamiltonians. It would further be desirable to identify in
detail the physical conditions under which the Pauli subgroup model is
relevant for current proposals for quantum computers. An important example
we have discussed is the dipolar-coupling induced decoherence in NMR.

\section{Acknowledgments}

This material is based upon work supported by the U.S. Army Research Office
under contract/grant number DAAG55-98-1-0371, and in part by NSF CHE-9616615.


\begin{thebibliography}{10}

\bibitem{Lo:book}
{H.K. Lo, S. Popescu and T.P. Spiller}, {\em Introduction to Quantum
  Computation and Information} (World Scientific, Singapore, 1999).

\bibitem{Williams:book98}
C. Williams and S. Clearwater, {\em Explorations in Quantum Computing}
  (Springer-Verlag, New York, 1998).

\bibitem{Duan:98}
{L.-M Duan and G.-C. Guo}, Phys. Rev. A {\bf 57},  737  (1998).

\bibitem{Duan:98c}
L. Duan and G. Guo, Phys. Rev. A {\bf 58},  3491  (1998), \uppercase{L}ANL
  Report No. quant-ph/9804014.

\bibitem{Zanardi:97c}
{P. Zanardi and M. Rasetti}, Phys. Rev. Lett. {\bf 79},  3306  (1997),
  \uppercase{L}ANL Report No. quant-ph/9705044.

\bibitem{Zanardi:97a}
{P. Zanardi and M. Rasetti}, Mod. Phys. Lett. B {\bf 11},  1085  (1997),
  \uppercase{L}ANL Report No. quant-ph/9710041.

\bibitem{Zanardi:98a}
P. Zanardi, Phys. Rev. A {\bf 57},  3276  (1998), \uppercase{L}ANL Report No.
  quant-ph/9705045.

\bibitem{Lidar:PRL98}
{D.A. Lidar, I.L. Chuang and K.B. Whaley}, Phys. Rev. Lett. {\bf 81},  2594
  (1998), \uppercase{L}ANL Report No. quant-ph/9807004.

\bibitem{Lidar:PRL99}
{D.A. Lidar, D. Bacon and K.B. Whaley}, Phys. Rev. Lett. {\bf 82},  4556
  (1999), \uppercase{L}ANL Report No. quant-ph/9809081.

\bibitem{Bacon:99}
{D. Bacon, D.A. Lidar and K.B. Whaley}, Phys. Rev. A {\bf 60},  1944  (1999),
  \uppercase{L}ANL Report No. quant-ph/9902041.

\bibitem{Shor:95}
{P.W. Shor}, Phys. Rev. A {\bf 52},  2493  (1995).

\bibitem{Steane:96a}
{A.M. Steane}, Phys. Rev. Lett. {\bf 77},  793  (1996).

\bibitem{Gottesman:97}
{D. Gottesman}, Phys. Rev. A {\bf 54},  1862  (1996), \uppercase{L}ANL Report
  No. quant-ph/9604038.

\bibitem{Knill:97b}
{E. Knill and R. Laflamme}, Phys. Rev. A {\bf 55},  900  (1997).

\bibitem{Viola:98}
L. Viola and S. Lloyd, Phys. Rev. A {\bf 58},  2733  (1998).

\bibitem{Viola:99}
{L. Viola, E. Knill and S. Lloyd}, Phys. Rev. Lett. {\bf 82},  2417  (1999).

\bibitem{Duan:98e}
L.-M. Duan and G. Guo, Pulse controlled noise suppressed quantum computation,
  \uppercase{L}ANL Report No. quant-ph/9807072.

\bibitem{Zanardi:98b}
P. Zanardi, Phys. Lett. A {\bf 258},  77  (1999), \uppercase{L}ANL Report No.
  quant-ph/9809064.

\bibitem{Bacon:99a}
{D. Bacon, J. Kempe, D.A. Lidar and K.B. Whaley}, {Universal Fault-Tolerant
  Computation on Decoherence-Free Subspaces}, submitted to Phys. Rev. Lett.
  Available as \uppercase{L}ANL Report No. quant-ph/9909058.

\bibitem{Kempe:00}
{J. Kempe, D. Bacon, D.A. Lidar, and K.B. Whaley}, {Theory of Decoherence-Free,
  Fault-Tolerant, Universal Quantum Computation}, submitted to Phys. Rev. A.
  Available as \uppercase{L}ANL Report No. quant-ph/0004064.

\bibitem{Lidar:PRA99Exchange}
{D.A. Lidar, D. Bacon, J. Kempe and K.B. Whaley}, Phys. Rev. A {\bf 61},
  052307  (2000), \uppercase{L}ANL Report No. quant-ph/9907096.

\bibitem{Slichter:96}
C. Slichter, {\em Principles of Magnetic Resonance}, No.~1 in {\em Springer
  Series in Solid-State Sciences} (Springer, Berlin, 1996).

\bibitem{Chemla:98}
{D.S. Chemla},  in {\em {Nonlinear Optics in Semiconductors}}, edited by {R.K.
  Willardson and A.C. Beers} ({Academic Press}, {New York}, 1998).

\bibitem{March:85vol2}
{W. Jones and N. March}, {\em {Theoretical Solid State Physics}} ({Dover}, {New
  York}, 1985), Vol.~2.

\bibitem{Lidar:00b}
{D.A. Lidar, D. Bacon, J. Kempe, and K.B. Whaley}, \uppercase{D}ecoherence-Free
  Subspaces for Multiple-Qubit Errors: (II) Universal, Fault-Tolerant Quantum
  Computation, submitted to Phys. Rev. A. Available as \uppercase{L}ANL Report
  No. quant-ph/0007013.

\bibitem{Leggett:87}
{A.J. Leggett, S. Chakravarty, A.T. Dorsey, M.A.P. Fisher, A. Garg and W.
  Zwerger}, Rev. Mod. Phys. {\bf 59},  1  (1987).

\bibitem{DiVincenzo:95}
{D.P. DiVincenzo}, Phys. Rev. A {\bf 51},  1015  (1995).

\bibitem{Lloyd:95}
{S. Lloyd}, Phys. Rev. Lett. {\bf 75},  346  (1995).

\bibitem{Barenco:95a}
{A. Barenco, C.H. Bennett, R. Cleve, D.P. DiVincenzo, N. Margolus, P. Shor, T.
  Sleator, J. Smolin and H. Weinfurter}, Phys. Rev. A {\bf 52},  3457  (1995).

\bibitem{Bhatia}
{R. Bhatia}, {\em {Matrix Analysis}}, No.~169 in {\em {Graduate Texts in
  Mathematics}} (Springer-Verlag, {New York}, 1997).

\bibitem{Gershenfeld:97}
{N. Gershenfeld and I.L. Chuang}, Science {\bf 275},  350  (1997).

\bibitem{Preskill:99}
{J. Preskill},  in {\em {Introduction to Quantum Computation and Information}},
  edited by {H.K. Lo, S. Popescu and T.P. Spiller} ({World Scientific},
  Singapore, 1999), \uppercase{L}ANL Report No. quant-ph/9712048.

\bibitem{Boykin:99}
{P. Boykin, T. Mor, M. Pulver, V. Roychowdhury, and F. Vatan}, {On Universal
  and Fault-Tolerant Quantum Computing}, \uppercase{L}ANL Report No.
  quant-ph/9906054.

\bibitem{Chuang:97c}
{I.L. Chuang and M.A. Nielsen}, J. Mod. Optics {\bf 44},  2455  (1997).

\bibitem{Gardiner:book}
C. Gardiner, {\em Quantum Noise}, Vol.~56 of {\em Springer Series in
  Synergetics} (Springer-Verlag, Berlin, 1991).

\bibitem{Kraus:83}
{K. Kraus}, {\em {States, Effects and Operations}}, {\em {Fundamental Notions
  of Quantum Theory}} ({Academic}, Berlin, 1983).

\bibitem{Schumacher:96a}
{B. Schumacher}, Phys. Rev. A {\bf 54},  2614  (1996).

\bibitem{Peres:99}
{A. Peres}, {Classical interventions in quantum systems. I. The measuring
  process}, \uppercase{L}ANL Report No. quant-ph/9906023.

\bibitem{March:book}
{N. March, W.H. Young and S. Sampanthar}, {\em {The Many-Body Problem in
  Quantum Mechanics}} ({Dover}, {New York}, 1995).

\bibitem{Duan:99}
L.-M. Duan and G.-C. Guo, Phys. Rev. A {\bf 59},  4058  (1999).

\bibitem{Bandyopadhyay:98}
{S. Bandyopadhyay, A. Balandin, V.P. Roychowdhury and F. Vatan}, Superlattices
  and Microstructures {\bf 23},  445  (1998), further detail in LANL Report No.
  quant-ph/9704019.

\bibitem{Schensted}
I.~V. Schensted, {\em A Course on the Application of Group Theory to Quantum
  Mechanics} (NEO Press, Peaks Island, Maine, 1976).

\bibitem{Nielsen:97}
{M.A. Nielsen and C.M. Caves}, Phys. Rev. A {\bf 55},  2547  (1997).

\bibitem{Cornwell:97}
{J.F. Cornwell}, {\em {Group Theory in Physics: An Introduction}} ({Academic
  Press}, San Diego, 1997).

\bibitem{Jaksch:99}
{D. Jaksch, H.-J. Briegel, J.I. Cirac, C.W. Gardiner and P. Zoller}, Phys. Rev.
  Lett. {\bf 82},  1975  (1999).

\bibitem{Ruskai:99}
{M.B. Ruskai}, {Pauli Exchange Errors in Quantum Computation}, Phys. Rev.
  Lett. {\bf 85},  194  (2000). \uppercase{L}ANL  Report No. quant-ph/9906114.

\end{thebibliography}

\newpage

\appendix

\section{The Pauli Group}

\label{app1}

The Pauli matrices are:

\begin{eqnarray}
\sigma _{0}\equiv I=\left( 
\begin{array}{cc}
1 & 0 \\ 
0 & 1
\end{array}
\right) \qquad \sigma _{x}=\left( 
\begin{array}{cc}
0 & 1 \\ 
1 & 0
\end{array}
\right) \qquad \sigma _{y}=\left( 
\begin{array}{cc}
0 & -i \\ 
i & 0
\end{array}
\right) \qquad \sigma _{z}=\left( 
\begin{array}{cc}
1 & 0 \\ 
0 & -1
\end{array}
\right) .
\end{eqnarray}

They have the following properties: 
\begin{eqnarray}
\sigma _{\alpha }^{2} &=&I\qquad \alpha =0,x,y,z  \nonumber \\
\lbrack \sigma _{\alpha },\sigma _{\beta } \rbrack &=&2i\varepsilon _{\alpha
\beta \gamma }\sigma _{\gamma }  \nonumber \\
\{\sigma _{\alpha },\sigma _{\beta }\} &=&2\delta _{\alpha \beta }I 
\nonumber \\
\sigma _{\alpha }\sigma _{\beta } &=&i\varepsilon _{\alpha \beta \gamma
}\sigma _{\gamma }+\delta _{\alpha \beta }I.  \nonumber \\
{\rm Tr}(\sigma _{\alpha }) &=&0\qquad \alpha =x,y,z
\end{eqnarray}

The {\em Pauli group} of order $K$ is the set of all $4^{K+1}$ possible
tensor products of $K$ of the Pauli matrices and $\pm ,\pm i$: 
\begin{eqnarray}
P_{K}=\pm ,\pm i\left\{ \bigotimes_{k=1}^{K}\sigma _{\alpha ,k}\right\}
_{\alpha }.
\end{eqnarray}

Some of its useful properties are

\begin{itemize}
\item  Let $p_{1},p_{2}\in P_{K}$. Since either $[\sigma _{\alpha ,k},\sigma
_{\beta ,k}]=0$ or $\{\sigma _{\alpha ,k},\sigma _{\beta ,k}\}=0$ it follows
that 
\begin{eqnarray}
\text{either }[p_{1},p_{2}]=0\text{ or }\{p_{1},p_{2}\}=0.
\end{eqnarray}

\item  Since $\sigma _{\alpha }$ are all unitary, so are all $p\in P_{K}$.

\item  Since $\sigma _{\alpha }$ are all Hermitian but we allow for $\pm i$
factors, $p\in P_{K}$ is either Hermitian or anti-Hermitian. Thus if $p\in
P_{K}$ then $p^{\dagger }\in P_{K}$.

\item  Since ${\rm Tr}(A\otimes B)={\rm Tr}A\times {\rm Tr}B$, the only
elements in $P_{K}$ which are not traceless are the four $\pm ,\pm i$
multiples of the identity, and each has trace $2^{K}$.
\end{itemize}

\section{Examples of Non-Generic Decoherence-Free Subspaces}

\label{app2}

We will show here an example of a DFS that arises out of a two-dimensional
irrep of a {\em non-Abelian} subgroup, in the ``non-generic'' case.

Let us consider the non-Abelian 8-element subgroup $Q_{8}$ $=\{\pm III,\pm
XXI,\pm IZZ,\pm iXYZ\}$. In this standard representation it is reducible and
splits into 4 copies of a two-dimensional irreducible representation of $
Q_{8}$. Since there is just one irrep, we drop the irrep-index $k$ on $
\Gamma ^{k}$ etc. The two-dimensional representation of $Q_{8}$ is the
following:

\begin{eqnarray}
\Gamma (\pm III) &=&\pm \left( 
\begin{array}{cc}
1 & 0 \\ 
0 & 1
\end{array}
\right) \qquad \Gamma (\pm XXI)=\pm \left( 
\begin{array}{cc}
0 & 1 \\ 
1 & 0
\end{array}
\right) \qquad  \nonumber \\
\Gamma (\pm IZZ) &=&\pm \left( 
\begin{array}{cc}
1 & 0 \\ 
0 & -1
\end{array}
\right) \qquad \Gamma (\pm iXYZ)=\pm \left( 
\begin{array}{cc}
0 & -1 \\ 
1 & 0
\end{array}
\right) .
\end{eqnarray}
The 8-dimensional Hilbert space is split into 4 irreducible subspaces $V^{z}$
(corresponding to the 4 copies of $\Gamma $) spanned by 
\begin{eqnarray}
V^{1} &=&(|\psi _{0}^{1}\rangle ,|\psi _{1}^{1}\rangle )=(|000\rangle
,|110\rangle )  \nonumber \\
V^{2} &=&(|\psi _{0}^{2}\rangle ,|\psi _{1}^{2}\rangle )=(|111\rangle
,|001\rangle )  \nonumber \\
V^{3} &=&(|\psi _{0}^{3}\rangle ,|\psi _{1}^{3}\rangle )=(|100\rangle
,|010\rangle )  \nonumber \\
V^{4} &=&(|\psi _{0}^{4}\rangle ,|\psi _{1}^{4}\rangle )=(|011\rangle
,|101\rangle ).
\end{eqnarray}
On each of these two-dimensional subspaces the group acts like $\Gamma $. A
codeword in the DFS can be expanded as $|\tilde{j}\rangle
=\sum_{z=1}^{4}\sum_{\mu =0}^{1}\theta _{z,\mu }^{j}|\psi _{\mu }^{i}\rangle 
$.\footnote{
Note that our indices here differ somewhat from the notations in Sec. \ref
{reptheory}, because there we considered either one-dimensional irreps, or
mostly avoided explicitly indicating superpositions between copies of a
given irrep.} Let us take as our code just the first basis-vector of each
irreducible subspace, i.e., 
\begin{eqnarray}
{\cal C}=\{|\tilde{1}\rangle ,|\tilde{2}\rangle ,|\tilde{3}\rangle ,|\tilde{
4 }\rangle \}\equiv \{|\psi _{0}^{z}\rangle :z=1\ldots 4\}=\{|000\rangle
,|111\rangle ,|100\rangle ,|011\rangle \}.
\end{eqnarray}
Denoting the vector of coefficients as $\overrightarrow{\theta _{z}^{j}}
=(\theta _{z,0}^{j},\theta _{z,1}^{j})$, this means that $\overrightarrow{
\theta _{z}^{z}}={(1,0)}$ and $\overrightarrow{\theta _{z}^{j\neq z}}$ $={\
(0,0)}$. In this case we can show that there are Kraus operators ${\bf A}
_{d} $ that satisfy the DFS-condition on the code, by searching for matrices 
${\cal A}_{d}$ which have $\overrightarrow{\theta _{z}^{j}}$ as
eigenvectors. An example is: 
\begin{eqnarray}
{\cal A}_{1}=\left( 
\begin{array}{cc}
c_{1} & d_{1} \\ 
0 & e_{1}
\end{array}
\right) \qquad {\cal A}_{2}=\left( 
\begin{array}{cc}
c_{2} & d_{2} \\ 
0 & e_{2}
\end{array}
\right) ,
\end{eqnarray}
with the conditions $c_{1}^{\ast }d_{1}+c_{2}^{\ast }d_{2}=0$, $
|c_{1}|^{2}+|c_{2}|^{2}=1$ and $
|d_{1}|^{2}+|d_{2}|^{2}+|e_{1}|^{2}+|e_{2}|^{2}=1$ for normalization [Eq.
(~ref{eq:OSRnorm})]. The corresponding Kraus-operators are: 
\begin{eqnarray}
{\bf A}_{1} &=&\frac{c_{1}+e_{1}}{2}III+\frac{d_{1}}{2}XXI+\frac{c_{1}-e_{1} 
}{2}IZZ+\frac{d_{1}}{2}iXYZ \\
{\bf A}_{2} &=&\frac{c_{2}+e_{2}}{2}III+\frac{d_{2}}{2}XXI+\frac{c_{2}-e_{2} 
}{2}IZZ+\frac{d_{2}}{2}iXYZ.  \nonumber
\end{eqnarray}
The code ${\cal C}$ is a DFS. It is the particular equality (i.e., the
``conspiring'', non-generic or accidental relationship) between the
coefficients of the $XXI$ and $XYZ$ terms that is responsible for the
existence of this DFS.

\end{document}